\DocumentMetadata{}
\documentclass[sigconf,review]{acmart}

\makeatletter
\def\mdseries@tt{m}
\makeatother

\pdfoutput=1

\usepackage{amsmath}
\usepackage{amsfonts}
\usepackage{graphicx}
\usepackage{booktabs}
\usepackage{array}
\usepackage{url}
\usepackage{listings}
\usepackage{xcolor}
\usepackage{caption}
\usepackage{subcaption}
\usepackage{float}
\usepackage{geometry}
\usepackage{algorithm}
\usepackage{algorithmic}
\usepackage{multirow}
\usepackage{mdframed}
\usepackage{xstring}

\usepackage{tikz}
\definecolor{GoodGreen}{RGB}{0,139,69}
\definecolor{BadRed}{RGB}{204,0,0}

\newcommand{\impact}[2]{%
  \IfStrEq{#2}{red}{%
    \textcolor{BadRed}{\textbf{#1}}%
  }{%
    \IfStrEq{#2}{green}{%
      \textcolor{GoodGreen}{\textbf{#1}}%
    }{%
      \textbf{#1}%
    }%
  }%
}

\begin{document}
\title[Debug2Fix]{Debug2Fix: Can Interactive Debugging Help Coding Agents Fix More Bugs?}

\author{Spandan Garg}
\authornote{Corresponding author.}
\email{spgarg@microsoft.com}
\affiliation{%
  \institution{Microsoft}
  \country{USA}
}
\author{Yufan Huang}
\email{yufanhuang@microsoft.com}
\affiliation{%
  \institution{Microsoft}
  \country{USA}
}

\date{}

\begin{abstract}
While significant progress has been made in automating various aspects of software development through coding agents, there is still significant room for improvement in their bug fixing capabilities. Debugging and investigation of runtime behavior remains largely a manual, developer-driven process. Popular coding agents typically rely on either static analysis of the code or iterative test-fix cycles, which is akin to trial and error debugging. We believe that there is a wealth of rich runtime information that developers routinely access while debugging code, which agents are currently deprived of due to design limitations. Despite how prevalent debuggers are in modern IDEs and command-line tools, they have surprisingly not made their way into coding agents. In this work, we introduce Debug2Fix, a novel framework that incorporates interactive debugging as a core component of a software engineering agent via a subagent architecture. We incorporate debuggers for Java and Python into our agent framework and evaluate against GitBug-Java and SWE-Bench-Live and achieve $>$20\% improvement in performance compared to the baseline for certain models. Furthermore, using our framework, we're able to make weaker models like GPT-5 and Claude Haiku 4.5 match or exceed the performances of stronger models like Claude Sonnet 4.5, showing that better tool design is often just as important as switching to a more expensive model. Finally, we conduct systematic ablations demonstrating the importance of both the subagent architecture and debugger integration.
\end{abstract}

\maketitle

\thispagestyle{empty}
\pagestyle{plain}

\section{Introduction}

Coding agents~\cite{claude_code_2024, vscode_2025, wang2024opendevinopenplatformai} demonstrate impressive capabilities in a variety of software engineering tasks. Bug-fixing is among the most common tasks developers perform with agents~\cite{saving}, yet there remains significant room for improvement~\cite{swebench_illusion, swepro, zan2025multiswebenchmultilingualbenchmarkissue}. Taking a closer look at how agents approach bugs sheds some light on some of the limitations in their bug-fixing capabilities. When faced with bugs or failing tests, coding agents follow one of two strategies: making changes based on pure static code analysis, or entering extended print-debugging sessions, both of which requires the agent to guess the underlying state of the program during execution. In the latter, the agent cycles between examining the program output i.e. reading error messages, stack traces, console logs, etc. to guess the underlying behavior and then making changes based on its interpretation of the output~\cite{selfdebug, shinn2024reflexion, selfrefine}. If the error persists, the agent continues to iterate through the debug and fix steps until the error goes away or it gives up. While this approach works for simpler bugs, the cycle of print-debugging and fixing is not only slow because it depends on repeatedly executing the program with small incremental changes that slowly approach the correct fix, but also unreliable as the agent is guessing runtime behavior rather than observing it directly. Without having access to the actual program state, the agent may make its decisions based on faulty hypotheses and make incorrect changes, which can even seep through human-written test suites~\cite{Wang2025AreI}.

We argue that this kind of print-debugging that agents primarily rely on, forms a major inefficiency in the debugging capabilities of today's agents. In contrast to this, expert human developers have long relied on debuggers to diagnose bugs effectively. Debuggers allow developers to pause the program at user-specified locations known as breakpoints, and inspect the values of variables, evaluate arbitrary expressions of code, examine the ongoing call-stack and step through code line by line. This is significantly more precise than inferring program behavior based on simply reading code or print output, where one wrong guess can compound and derail a whole debug session. Furthermore, this is also more efficient because converging on the right set of breakpoints requires fewer iterations than incrementally building an understanding of code based on print outputs alone.

Despite the prevalence of debuggers in modern IDEs and command-line (CLI) tools, they have surprisingly not made their way into coding agents. This is likely due to the fact that debuggers were designed with human-interaction in mind. They require careful orchestration of commands to interact with the program state. Command-line debugger commands are verbose by design as they're designed to provide as much diagnostic information as possible. Finally, debuggers involve asynchronous events and timing, which is not easy for an agent to manage programmatically. As a consequence, naively exposing a debugger to the agent can lead to brittle interactions and failures. Due to these limitations, an effective schema for debuggers is a non-trivial task. However, it's also possible that one might go through the exercise of creating a perfect suite of tools only for the agent to never use it~\cite{livemcp}. In our experiments, providing debug tools directly to the main agent resulted in minimal usage. Despite all these difficulties, we believe that agents not leveraging debuggers represents a significant missed opportunity. 


\begin{figure*}
    \centering
    \includegraphics[width=0.9\linewidth]{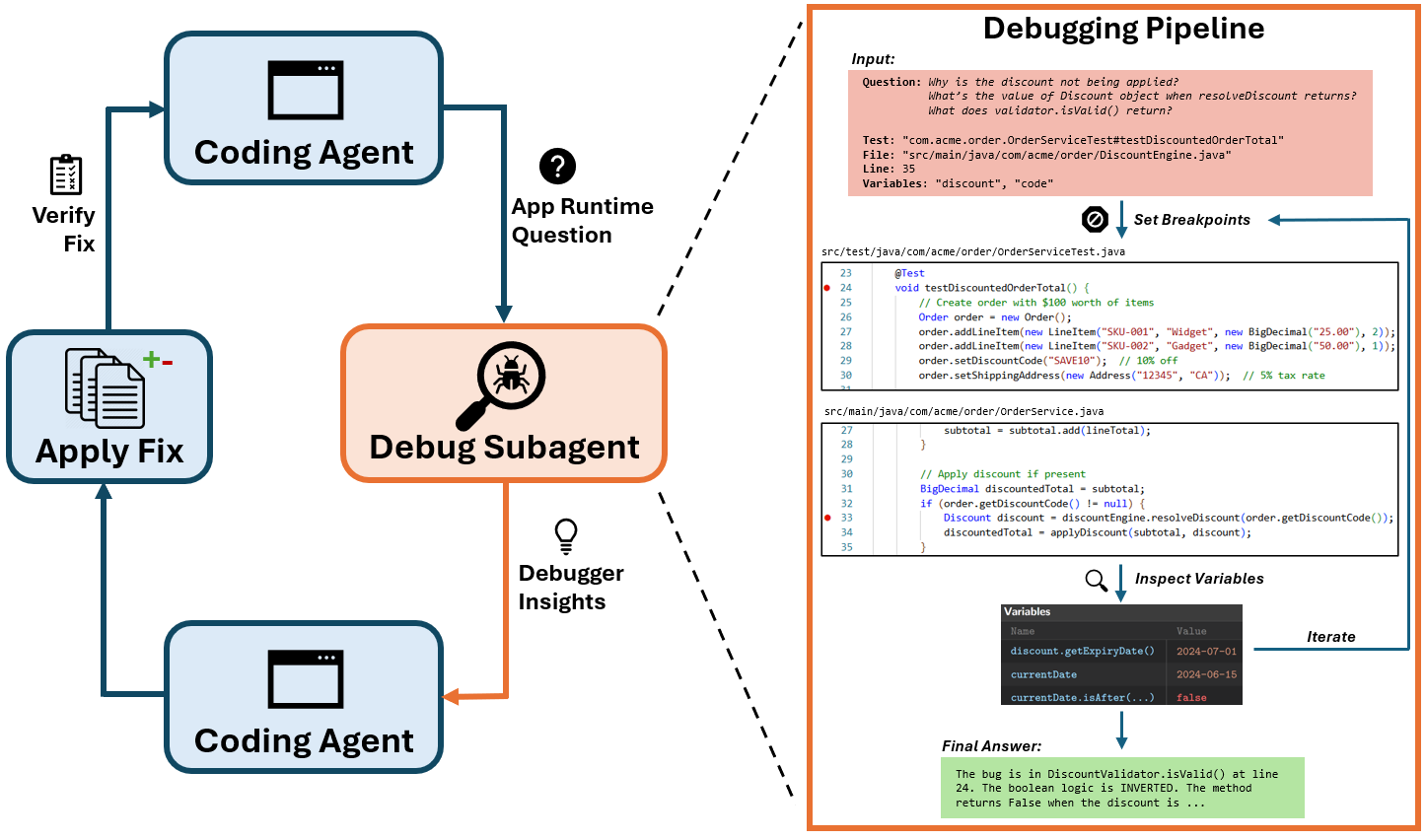}
    \caption{A high-level view of the overall Debug2Fix pipeline with the Debug Subagent. We can see that the main agent's is to loop between querying the Debug Subagent, followed by making fixes based on the learned insights from runtime behavior. Internally, the Debug Subagent works by going through a cycle of setting breakpoints, stepping through code and inspecting variables / expressions until it has the answer to main agent's query or runs out of turns.}
    \label{fig:architecture}
\end{figure*}

To address all these gaps we introduce \textbf{Debug2Fix}, the first specialized agent framework that incorporates debugging as a core component of the agent framework. It works by incorporating debuggers into coding agents via a novel tool design and crucially, a subagent architecture. Rather than exposing the debugger tools to the main agent directly, where they most likely go unused, we expose a unified Debug Subagent to the main agent, which is strongly instructed to use it.
This subagent design encapsulates all the complexities associated with debuggers behind a simple high-level interface for the main agent to use. During inference, the agent offloads debugging tasks to the subagent, which then handles the debugger orchestration via its set of tools and returns a concise answer with all the findings. Our contributions in this work are as follows:

\begin{itemize}
    \item \textbf{Debug2Fix}: We present the first integration of interactive debugging into a coding agent via the use of a subagent and a novel tool schema, enabling complex runtime debugging capabilities that complement static code analysis. Through our qualitative analysis, we show that the Debug Subagent follows similar workflows as an expert human developer.
    \item \textbf{Empirical Evaluation \& Ablation Study}: We conduct an extensive evaluation of our approach by incorporating the Java Debugger (JDB)~\cite{jdb} and Python Debugger (PDB)~\cite{pdb} into a coding agent and evaluate on bug fixing benchmarks like GitBug-Java~\cite{gitbug} and SWE-Bench-Live~\cite{swelive}. We show that adopting Debug2Fix framework improves performance over vanilla agent by $>$20\% relative to baseline numbers in some cases. We also conduct ablation experiments showing that both our tool design and subagent architecture are essential.
\end{itemize}



\section{Background and Related Work} 
We build upon a rich foundation of research in Software Engineering and Agentic AI. Our work bridges a crucial gap in Software Engineering Agents and the typical developer workflow for fixing bugs.

\subsection{Automated Debugging}
Debugging is a core skill in software development. However it remains largely a manual and human-driven process. There have been works that try to automate debugging. AutoSD~\cite{autosd} prompts LLMs to automatically generate hypotheses and uses debuggers to interact with buggy code. ChatDBG~\cite{chatdbg} augments traditional debuggers into an AI-powered debugging assistant. While still a human-driven process, it integrates LLMs into debuggers to enhance the user-friendliness of conventional debuggers and allowing humans to have a dialogue with the debugger and pose complex questions. Zhong et al.~\cite{debughuman} propose an LLM Debugger (LDB) that allows LLMs to refine their own generated programs with runtime information. It segments programs into blocks and tracks intermediate values after each block. debug-gym~\cite{debuggym} introduces a text-based environment that exposes pdb directly to a single LLM agent.

While these approaches demonstrate the value of runtime information and various forms of debugging, they differ from our work in key ways. Our work is specifically designed to introduce debuggers via subagent architecture into automated coding agents that work on the entire repository. It encapsulates debugger complexity behind a high-level interface so that agents can utilize it effectively.

\subsection{Software Engineering Agents}
There have been significant advancements in the field of SE Agents. Starting with SWE-Agent~\cite{yang2024sweagent}, which proposed key design principles that demonstrated high-performance on repository-level tasks. Since then agent systems like OpenHands~\cite{wang2024opendevinopenplatformai}, Claude Code~\cite{claude_code_2024}, Copilot CLI~\cite{copilot_agent_2024}, VSCode Agent~\cite{vscode_2025}, Windsurf~\cite{windsurf_2024} have seen widespread adoption into the developer workflow and demonstrate impressive performance on various software engineering tasks such as test generation, bug fixing, code search, etc.

While existing agent frameworks have shown impressive capabilities on various benchmarks, gaps still remain when it comes to harder bug-fixing tasks. These agents leverage suboptimal techniques like iterative print-debugging cycles by inserting logging statements into code or making educated guesses based on static analysis of code. Our work addresses these core limitations with agents by incorporating interactive debugging into coding agents as a dedicated subagent.

\subsection{Multi-Agent Architectures}
The use of specialized agents has proven to be an effective pattern in AI Agent Systems. Rather than expecting a single agent / model to have all capabilities, multi-agent systems decompose the high-level problem and delegate responsibilities to smaller agents or subagents. Many works have improved problem-solving abilities of LLMs by integrating discussions among multiple agents. In their survey~\cite{mamo}, He et al. systematically reviewed the landscape of LLM-based multi-agent systems for software engineering highlighting the current capabilities and limitations of these approaches. MASAI~\cite{masai} proposes a modular architecture for software engineering agents where subagents are instantiated with well-defined objectives and strategies, achieving competitive performance on coding benchmarks. AutoDev~\cite{tufano2024autodevautomatedaidrivendevelopment} uses multiple autonomous AI Agents to achieve user defined objectives. MapCoder~\cite{mapcoder} is another such framework that consists of several LLM agents, which is designed to simulate the stages of the developer cycle. FixAgent~\cite{fixagent} applies multi-agent synergy to debugging, using specialized agents inspired by rubber-duck debugging.

Our Debug Subagent follows a modular multi-agent pattern, providing the main agent with a high-level interface for debugging queries while encapsulating the complexity of low-level debugger interactions. Unlike prior works that decompose tasks into code search, localization, repair phases, we introduce a specialized capability that complements static analysis. Through ablation experiments, we show that this architectural choice is essential i.e. when debugging tools are exposed to the main agent directly, they go largely unused. The subagent design is pivotal in making this otherwise low-level capability into a simple interface the main agent readily utilizes.

\begin{figure*}[t]
\centering

\begin{minipage}[t]{0.97\textwidth}
\textbf{\small Issue Report:}\hfill\texttt{\footnotesize joke2k/faker\#2154}
\begin{mdframed}[style=issuebox]
\scriptsize
For me this started breaking when using v34.0.1
\begin{lstlisting}[language=Python, basicstyle=\ttfamily\tiny, breaklines=true]
self = <faker.providers.date_time.en_US.Provider object at 0x7f55c5ecf210>
start_datetime = 0, end_datetime = -604800

    def _rand_seconds(self, start_datetime: int, end_datetime: int) -> float:
        if start_datetime > end_datetime:
>           raise ValueError("empty range for _rand_seconds: ...")
E           ValueError: empty range for _rand_seconds:
E             start datetime must be before than end datetime
\end{lstlisting}
\vspace{-0.5em}
Not sure if my code is wrong or something, but all was working fine before.

This is what is failing I think:\\
\texttt{deadline = factory.Faker("date\_time", end\_datetime="-1w", tzinfo=timezone.get\_current\_timezone())}
\end{mdframed}
\end{minipage}

\vspace{0.8em}

\begin{minipage}[t]{0.98\textwidth}
\hspace*{1.54cm}\includegraphics[width=0.8\textwidth]{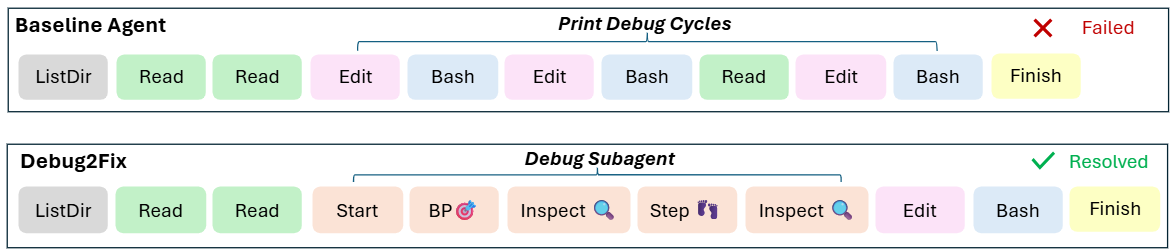}
\end{minipage}

\caption{A bug from a popular open-source Python repository on GitHub. We see very different trajectories taken by the Baseline Agent and Debug2Fix. In the baseline, we see the agent doing repeated print-debug cycles and arriving at the wrong fix due to not being able to find the root cause of the issue, which is situated deep within the repo. With Debug2Fix, the agent uses the Debug Subagent which is able to find the root cause immediately using a debugger. This results in the agent arriving at the correct fix.}
\label{fig:motivating-example}
\end{figure*}

\section{Motivating Example}
In this section, we provide an overview of Debug2Fix with a motivating example. Figure \ref{fig:motivating-example} shows an example of a bug from a popular open-source python library where the user is trying to generate a random Unix timestamp from up to a week ago (\texttt{-1w} in code). The library method works by parsing the \texttt{end\_datetime} string into a timestamp by computing a time delta of negative seven days compared to the output of the \texttt{\_safe\_now()} method, which is supposed to return the current timestamp. However, buried deep in the class hierarchy, its fallback is set to \texttt{datetime(1970, 1, 1)} in one of the code paths. Since this is `Unix Time \texttt{0}', taking a negative timestamp from it results in a crash.

We compare the trajectories of both the baseline agent and our Debug2Fix agent when tasked with solving this problem given the corresponding repo. Looking at the high-level agent trajectories in the figure, we can see that the baseline agent engages in repeated edit and bash calls to do print-debugging, until it believes issue is fixed. Due to how deep in the codebase the issue is, the agent ends up having to do a lot of print debugging as it struggles to find where the bug is coming from. In this case, it actually ends up re-writing the unit test associated with this problem and submitting the wrong fix. Debug2Fix solves this problem with relative ease. Rather than guessing at the program's runtime behavior through print statements, Debug2Fix delegates the problem of finding the root cause to the Debug Subagent as seen in the figure. The subagent sets a breakpoint directly in the test script and inspects the local variables, while stepping through the code. Looking at the return value of \texttt{\_safe\_now()}, which returns \texttt{datetime(1970, 1, 1)} instead of the current time immediately reveals the root cause of the issue.

This set of contrasting trajectories was one of many observed across our benchmark evaluations.


\section{Methodology}

In this section, we explain the approach behind Debug2Fix in detail and how we incorporate it into an existing coding agent. Figure \ref{fig:architecture} shows a high-level view of the overall approach. We first describe the schema and underlying architecture of the Debug Subagent. Let's begin by describing what a subagent is:

\textbf{Subagent}: A subagent is a secondary LLM-based agent that is invoked by the primary (or "main") agent to handle a specialized sub-task. Unlike the main agent, which is meant to solve a broad set of tasks, the subagent solves a smaller, often simpler, set of tasks using a specialized list of tools. Just like the main agent, the subagent has its own system prompt, context window and tool-set that it uses to achieve the goal delegated to it by the main agent.

\subsection{Debug Subagent}
The Debug Subagent is one such specialized subagent designed to answer questions about runtime behavior. In our case, it's exposed to the main agent as a simple tool, which can be invoked with the following set of parameters and arguments:
\begin{itemize}
    \item \textbf{Runtime Question} (Required): The main agent describes what it wants to know about the program's runtime behavior. This is the only required field.
    \item \textbf{Test(s)} (Required): The main agent provides a script / test that fails.
    \item \textbf{Path} (Optional): This is a path to the code the main agent wants the subagent to exercise.
    \item \textbf{Lines} (Optional): These are the lines we want to set the initial breakpoints on.
    \item \textbf{Variable} (Optional): These are the variables the main agent has told us to pay particular attention to.
\end{itemize}

As it's output, the Debug Subagent returns a structured response containing a final answer to the question asked by the main agent and supporting evidence (observed variable values, stack traces, source locations, etc.).

This simple question-answer abstraction shields the main agent from the underlying complexity of the debugging orchestration. Rather than exposing low-level debugger commands to the main agent, this interface deliberately gives it the flexibility to ask questions in natural language while providing helpful metadata and the subagent to adapt accordingly.

\subsection{Debug Tools}
The Debug Subagent interacts with a set of tools that encapsulate the underlying debugger used for that language. These tools are designed to mirror the core actions a human developer would perform with a debugger, but in an LLM-friendly way. Below we describe each tool one by one.

\textbf{Debug Start Session}: This is the most critical tool that we add and gets called at the start of every Debug Subagent trajectory and performs the entire setup sequence atomically. It detects the underlying build system, builds the project, launches the test, waiting for a debug port to be available, attaching a debugger to it and setting any initial breakpoints. By default the tool also sets a breakpoint at the start of the test method provided by the main agent. Building all these steps into a single tool was a deliberate design choice. In our early experiments, exposing each step as a separate tool resulted in frequent failures, race conditions and timeouts. The atomic design is key to systematically eliminating all these issues.

\textbf{Debug Control}: Once the debug session is active and paused at a breakpoint, the subagent can use this tool to control the flow of execution. A debugger typically allows a user to take one of the following actions after a breakpoint is hit: continue, step over, step into or step out. These allow the user to let the execution continue to the next breakpoint, step over to the next line after executing the current one, step into the method being called on the current line or finally, exit the method to the calling method's context. Rather than one tool for each of these actions, we provide one central control tool that takes an enum for which action the LLM wants the subagent to take. Finally, the text output of this tool shows the surrounding context and the lines that have breakpoints.

\textbf{Debug Inspect}:  Once the execution is paused, this tool allows the subagent to query program state. It supports examining local variables, evaluating arbitrary expressions, viewing call stacks and inspecting fields within objects. This is where the subagent gathers most of the information it needs to answer the main agent question and the evidence needed.

\textbf{Debug Breakpoint}: This tool simply allows the subagent to set or remove any breakpoints. It also allows for listing all the breakpoints. For toggling breakpoints, we allow line and method breakpoints, which set the breakpoints on a specific line or the entrance of a method respectively. This is to allow the LLM the flexibility to set breakpoints based on method names, which are less error-prone than line numbers.

In addition to these tools, the Debug Subagent is also equipped with tools for file navigation, grep search and reading files to allow the subagent to gather context before setting breakpoints, deciding execution control, interpreting call stacks, etc. We leave out tools for file editing because the goal isn't to fix the issue only to identify the root cause. Finally, Figure \ref{fig:subagent_prompt} shows the system prompt we use for our subagent.

\begin{figure}[h]
\centering
\fbox{%
\begin{minipage}{0.9\columnwidth}
\scriptsize
\ttfamily
\setlength{\parskip}{1pt}
\noindent You are a Runtime Oracle - a debugging assistant that\\
answers specific questions about program execution.\\[4pt]
\textbf{Assumption}\\
The project is ALREADY BUILT. The main agent has compiled\\
the code before calling you. Do NOT attempt to build the\\
project yourself - go directly to debugging.\\[4pt]
\textbf{Your Role}\\
You answer questions about runtime behavior by:\\
1. Starting a debug session with initial breakpoints\\
2. Inspecting variables when breakpoints hit\\
3. Stepping through code as needed\\
4. Returning factual, verifiable answers\\[4pt]
\textbf{Question Types You Handle}\\
- Variable Inspection: "What is the value of X at line Y?"\\
- Reachability: "Does execution reach line Z during test T?"\\
- Condition Evaluation: "Why does condition X evaluate to true?"\\
- Exception Origin: "What causes the NullPointerException?"\\[4pt]
\textbf{Output Format (REQUIRED)}\\
Always end your investigation with a <debug\_answer> block:\\[2pt]
\phantom{xx}<debug\_answer>\\
\phantom{xx}**Question**: [The question you were asked]\\
\phantom{xx}**Answer**: [Direct, factual answer]\\
\phantom{xx}**Evidence**: [Variable values, stack frames observed]\\
\phantom{xx}**Location**: [File:line where you observed this]\\
\phantom{xx}</debug\_answer>\\[4pt]
\textbf{Tools Available}\\
- debug\_start\_session: Start session with test and breakpoints\\
- debug\_inspect: Inspect variables, evaluate expressions, view stack\\
- debug\_control: Step through code, continue, terminate\\
- debug\_breakpoint: Add or remove breakpoints\\
- read\_file: Read source code for context\\
\end{minipage}%
}
\caption{System prompt for the Debug Subagent. The prompt explains the role of the subagent to the LLM along with descriptions of the kinds of questions, tools available and the output format required.}
\label{fig:subagent_prompt}
\end{figure}

\begin{figure}[h]
\centering
\fbox{%
\begin{minipage}{0.87\columnwidth}
\footnotesize
\ttfamily
\setlength{\fboxsep}{0.8pt}
\setlength{\parskip}{0.8pt}
\noindent You are a highly sophisticated automated coding agent\\
with expert-level knowledge across many different\\
programming languages and frameworks.\\
\noindent The user will ask a question, or ask you to perform\\
a task, and it may require lots of research to answer\\
correctly. There is a selection of tools that let you\\
perform actions or retrieve helpful context...\\[6pt]
\colorbox{green!20}{\textbf{== Using debug\_subagent for Bug Fixing ==}}\\[2pt]
\colorbox{green!20}{You have access to `debug\_subagent' - a debugging}\\
\colorbox{green!20}{tool that can inspect runtime values, trace execution,}\\
\colorbox{green!20}{and help verify fixes. Use it to understand bugs}\\
\colorbox{green!20}{before making changes.}\\[4pt]
\colorbox{green!20}{\textbf{Recommended Workflow:}}\\[2pt]
\colorbox{green!20}{\textbf{Step 1: Build the project first}}\\
\colorbox{green!20}{\phantom{xx}mvn test-compile -q}\\[2pt]
\colorbox{green!20}{\textbf{Step 2: Understand the bug} (before making changes)}\\
\colorbox{green!20}{\phantom{xx}debug\_subagent(\{question: "What exception}\\
\colorbox{green!20}{\phantom{xx}occurs when running MyTest\#testMethod?",}\\
\colorbox{green!20}{\phantom{xx}test: "com.example.MyTest\#testMethod"\})}\\[2pt]
\colorbox{green!20}{\textbf{Step 3: Investigate root cause}}\\
\colorbox{green!20}{\phantom{xx}debug\_subagent(\{question: "What is the value}\\
\colorbox{green!20}{\phantom{xx}of [variable] at [location]?"\})}\\[2pt]
\colorbox{green!20}{\textbf{Step 4: Apply your fix}}\\[2pt]
\colorbox{green!20}{\textbf{Step 5: Verify the fix works}}\\
\colorbox{green!20}{\phantom{xx}debug\_subagent(\{question: "Does the test pass}\\
\colorbox{green!20}{\phantom{xx}now after my fix?"\})}\\[6pt]
\noindent If you aren't sure which tool is relevant, you can\\
call multiple tools. You can call tools repeatedly to\\
take actions or gather as much context as needed...\\
\end{minipage}%
}
\caption{Instructions added (green) to the main agent system prompt as part of the Debug2Fix framework. We inject a dedicated section that introduces the
Debug Subagent and provides a recommended workflow for bug-fixing tasks.}
\label{fig:prompt_diff}
\end{figure}

\subsection{Main Agent Integration}
Next, we describe how the Debug Subagent integrates with the main agent itself. We integrate the Debug Subagent as one of the tools available to the main agent, alongside tools like (Bash, Read, Grep, etc.). We also augment the main agent's system prompt (Figure \ref{fig:prompt_diff}) to describe when and how to invoke the Debug Subagent. This describes the kinds of tasks the main agent should delegate to the subagent: inspecting runtime values, root cause analysis, verifying fixes, etc. In our Qualitative Analysis, we see the Debug Subagent actually being used for these kinds of tasks. 

These prompt changes describe the high-level workflow shown in Figure \ref{fig:architecture} to the LLM, where we want the LLM to call the Debug Subagent before making any changes to the code instead of doing print-debugging. This mirrors the natural workflow of an experienced developer, who reaches for a debugger when faced with a complex issue where static code analysis isn't enough. In principle, these changes would suffice but we face an interesting problem, which we describe next.

\subsubsection{The Problem of Tool Under-utilization}
In some of our experiments, simply giving the main agent the subagent wasn't enough. The usage of the subagent varied significantly by model~\cite{livemcp} (Table~\ref{tab:consolidated_results_py}). To solve this problem, we employ a two-part strategy: 1) rather than exposing the debug tools directly, we expose the subagent and only the subagent to the main agent, 2) for bug-fixing tasks, we disable all file-editing tools until the debug tool has been called at least once. Following the workflow shown in Figure \ref{fig:architecture}, We believe that the main agent shouldn't have to modify any files until it has done root cause analysis with the Debug Subagent. It's allowed to navigate the codebase and view files, but it can only modify files after debugging the issue via the subagent. We justify this design choice with ablation experiments in our evaluation.

\section{Experimental Setup}
In this section, we talk about our experimental setup to evaluate the impact of adding a Debug Subagent following the methodology described earlier.

\subsection{Languages}
We implement Debug2Fix for two languages: Java and Python. For Java, we integrate the Java Debugger (JDB), a command line debugger that comes with the Java Development Kit (JDK). We add support for both Maven and Gradle build systems, which are automatically detected based on the files in the codebase. Since building the projects can be a long process, we use incremental build settings that are supported by these build systems.

For Python, we integrate the Python Debugger (PDB), which is part of the standard Python library and comes built in with the language. Unlike Java, Python does not require a separate compilation step and PDB can be invoked directly on any Python script. However, we added support for the more standard PyTest-based test execution as well.

\subsection{Benchmarks}
We use the following two benchmarks for our evaluation:
\begin{itemize}
    \item \textbf{SWE-Bench-Live}: A benchmark~\cite{swelive} extending the methodology of the original SWE-bench~\cite{swebench} with GitHub issues from 93 repos, filtered to have issues created after 2024 to minimize contamination. We use a subset of 400 python examples from their frozen Verified split.
    \item \textbf{GitBug-Java}: A Java benchmark~\cite{gitbug} from 55 notable open-source repos on GitHub. We use a subset of 186 examples for which we were able to successfully execute the provided docker images and evaluation harness.
\end{itemize}
Both benchmarks use a test-based verification i.e. a fix is considered correct if all relevant tests pass after the agent fix has been applied.

\subsection{Metrics}
We evaluate our approach using the following metrics:
\begin{itemize}
    \item \textbf{Pass Rate (\%)}: We observe the change in pass rate of the agent across different configurations.
    \item \textbf{Call Rate (\%)}: This is the \% of instances in the run where the Debug Subagent was invoked. This helps us understand if the subagent is actually being used, which is important for our ablation study.
    \item \textbf{Avg. Step Count}: This is the average number of steps taken by main agent and the Debug Subagent per instance.
    \item \textbf{Avg. Token Usage}: This is the sum of average input and output tokens used by an instance in a given run. This along with number of steps captures the computational cost and latency of running the agent with our subagent architecture.
\end{itemize}

Along with all these metrics, we also report the change a configuration has on that metric as a percentage of baseline value, shown in brackets after the actual metric.

\subsection{Models Configurations}
We evaluate Debug2Fix over 3 popular frontier LLMs from OpenAI and Anthropic: GPT-5, Claude Sonnet 4.5 and Claude Haiku 4.5. For all our models, we use the same model for both our main agent and Debug Subagent. We leave experiments with different model combinations (stronger model for main agent and smaller finetuned model for subagent) up to future experimentation.

\subsection{Ablations}
We ablate over different design choices via the following configurations:
\begin{itemize}
    \item \textbf{Baseline}: The agent without any modifications.
    \item \textbf{Debug Tools Only}: We add the debug tools defined in the Debug Subagent Tools section, but we expose them to the main agent directly. This is meant to evaluate the benefit of having a subagent architecture.
    \item \textbf{Debug2Fix}: We incorporate the Debug Subagent in the agent, as described in the Methodology.
    \item \textbf{Debug2Fix (w/ Tool Limit)}: We incorporate Debug Subagent and also disable Edit tools until the agent has called the Debug Subagent at least once.
\end{itemize}

For all the configurations that use the Debug Subagent, we limit each Debug Subagent trajectory to be at most 25 steps each. So, in other words, the Debug Subagent cannot take more than 25 steps at a time to investigate an issue. If the subagent hasn't finished by then, we prompt the agent to generate the final answer by injecting a user prompt into the trajectory and querying the LLM.

\begin{table*}[t]
\centering
\caption{Comparison of success rate for each configuration. We also show the breakdown of steps and tokens used by the main agent and the Debug Subagent within each of our configurations. For each number, we also show the relative improvement or decrease in performance over baseline.}
\label{tab:consolidated_results}
\begin{subtable}[t]{\textwidth}
\centering
\footnotesize
\caption{GitBug-Java}
\begin{tabular}{l|ll|ll|ll}
\toprule
\textbf{Metric} & \multicolumn{2}{l|}{\textbf{GPT-5}} & \multicolumn{2}{l|}{\textbf{Claude Haiku 4.5}} & \multicolumn{2}{l}{\textbf{Claude Sonnet 4.5}} \\
\midrule
\multicolumn{7}{c}{\textbf{Pass \& Call Rate (\%)}} \\
\midrule
& \tiny Pass \% & \tiny Call \% & \tiny Pass \% & \tiny Call \% & \tiny Pass \% & \tiny Call \% \\
Baseline & 60.2 & - & 71.0 & - & 75.7 & - \\
Debug Tools Only & 60.8 (\impact{+1.0\%}{green}) & - & 70.4 (\impact{-0.8\%}{red}) & - & 64.5 (\impact{-14.8\%}{red}) & - \\
Debug2Fix & 64.0 (\impact{+6.3\%}{green}) & 64.5 & 76.1 (\impact{+7.2\%}{green}) & 69.4 & 78.0 (\impact{+3.0\%}{green}) & 70.4 \\
Debug2Fix (w/ Tool Limit) & \textbf{73.1} (\impact{+21.8\%}{green}) & 99.5 & \textbf{82.3} (\impact{+15.9\%}{green}) & 98.9 & \textbf{85.5} (\impact{+12.9\%}{green}) & 98.9 \\
\midrule
\multicolumn{7}{c}{\textbf{Avg. Steps}} \\
\midrule
& \tiny Main & \tiny Sub & \tiny Main & \tiny Sub & \tiny Main & \tiny Sub \\
Baseline & 15.7 & - & 42.8 & - & 35.7 & - \\
Debug Tools Only & 27.9 (\impact{+77.7\%}{red}) & - & 44.4 (\impact{+3.7\%}{red}) & - & 33.3 (\impact{-6.7\%}{green}) & - \\
Debug2Fix & 19.1 (\impact{+21.7\%}{red}) & 33.7 & 43.2 (\impact{+0.9\%}{red}) & 52.1 & 34.4 (\impact{-3.6\%}{green}) & 39.8 \\
Debug2Fix (w/ Tool Limit) & 23.6 (\impact{+50.3\%}{red}) & 25.9 & 47.5 (\impact{+11.0\%}{red}) & 44.5 & 33.7 (\impact{-5.6\%}{green}) & 33.1 \\
\midrule
\multicolumn{7}{c}{\textbf{Avg. Tokens (Input + Output)}} \\
\midrule
& \tiny Main & \tiny Sub & \tiny Main & \tiny Sub & \tiny Main & \tiny Sub \\
Baseline & 347k & - & 1.55M & - & 1.22M & - \\
Debug Tools Only & 802k (\impact{+131\%}{red}) & - & 2.64M (\impact{+70.3\%}{red}) & - & 1.15M (\impact{-5.7\%}{green}) & - \\
Debug2Fix & 442k (\impact{+27.4\%}{red}) & 479k & 1.71M (\impact{+10.3\%}{red}) & 760k & 1.12M (\impact{-8.2\%}{green}) & 478k \\
Debug2Fix (w/ Tool Limit) & 645k (\impact{+85.9\%}{red}) & 350k & 1.75M (\impact{+12.9\%}{red}) & 619k & 978k (\impact{-19.8\%}{green}) & 396k \\
\bottomrule
\end{tabular}
\end{subtable}
\end{table*}

\section{Results}
We evaluate Debug2Fix across two benchmarks and three frontier LLMs from OpenAI and Anthropic. Table~\ref{tab:consolidated_results} summarizes our findings on GitBug-Java and SWE-Bench Live. For GitBug-Java, we conduct a detailed ablation study across all configurations described earlier. We then use SWE-Bench Live to validate whether our approach generalizes to Python. In this section, we drill deeper into the results of our runs.

\subsection{Main Results}
For GitBug-Java, Debug2Fix improves the pass rate across all three models. GPT-5 sees the largest gains, jumping $>$20\% over the baseline from 60.2\% to 73.1\%. Claude Sonnet also improves by $\sim$13\% over baseline, while Claude Haiku sees $\sim$16\% improvement.

We can also see that with the Debug2Fix configuration with the tool-limit in place, we see a $>$98\% call rate for Debug Subagent, showing that our strategy is effective in encouraging debugger usage. In contrast, when not used without the tool limit, we only see a 60-70\% call rate across the models. The Debug Subagent isn't provided to the Baseline and Debug Tools Only configuration, so the corresponding fields are blank.

\begin{table*}[t]
\centering
\footnotesize
\caption{Comparing the performances, token usages and steps used by Debug2Fix and Baseline agent on the Python subset of SWE-Bench Live dataset.}
\label{tab:consolidated_results_py}
\begin{subtable}[t]{\textwidth}
\centering
\caption{SWE-Bench Live (Python)}
\begin{tabular}{l|ll|ll|ll}
\toprule
\textbf{Metric} & \multicolumn{2}{c|}{\textbf{GPT-5}} & \multicolumn{2}{c|}{\textbf{Claude Haiku 4.5}} & \multicolumn{2}{c}{\textbf{Claude Sonnet 4.5}} \\
\midrule
\multicolumn{7}{c}{\textbf{Pass \& Call Rate (\%)}} \\
\midrule
& \tiny Pass \% & \tiny Call \% & \tiny Pass \% & \tiny Call \% & \tiny Pass \% & \tiny Call \% \\
Baseline & 31.2 & - & 34.3 & - & 39.6 & - \\
Debug2Fix & \textbf{36.2} (\impact{+16.0\%}{green}) & 61.4 & \textbf{38.5} (\impact{+12.2\%}{green}) & 33.2 & \textbf{40.4} (\impact{+2.0\%}{green}) & 8.1 \\
\midrule
\multicolumn{7}{c}{\textbf{Avg. Steps}} \\
\midrule
& \tiny Main & \tiny Sub & \tiny Main & \tiny Sub & \tiny Main & \tiny Sub \\
Baseline & 20.0 & - & 55.4 & - & 53.0 & - \\
Debug2Fix & 21.9 (\impact{+9.5\%}{red}) & 24.6 & 57.5 (\impact{+3.8\%}{red}) & 28.6 & 54.4 (\impact{+2.6\%}{red}) & 31.4 \\
\midrule
\multicolumn{7}{c}{\textbf{Avg. Tokens (Input + Output)}} \\
\midrule
& \tiny Main & \tiny Sub & \tiny Main & \tiny Sub & \tiny Main & \tiny Sub \\
Baseline & 550k & - & 1.97M & - & 1.96M & - \\
Debug2Fix & 630k (\impact{+14.5\%}{red}) & 350k & 2.08M (\impact{+5.6\%}{red}) & 371k & 1.97M (\impact{+0.7\%}{red}) & 281k \\
\bottomrule
\end{tabular}
\end{subtable}
\end{table*}

For SWE-Bench-Live, we evaluate the Debug2Fix agent without tool-limiting to understand the natural adoption patterns across models. GPT-5 shows the highest call rate of $>$60\% and has the corresponding largest improvement (16\%). Claude Haiku 4.5 demonstrates moderate usage (33.2\%) and seems modest gains as well of 12\%. Finally, we can see that when naturally prompted, Claude Sonnet 4.5 only calls the subagent in $\sim$8\% of the instances with minimal improvement. This makes sense because we haven't really changed the agent's workflow if the tool didn't trigger and it achieves a similar performance as the baseline. Further, this example also shows our approach generalizing to Python despite there being differences between the interfaces of JDB and PDB and how they're used.

\subsection{Ablation Analysis}
Our ablation study helps us isolate the contribution of the two key design choices: the subagent architecture and tool-limiting strategy.

\textbf{Exposing debug tools directly is ineffective and even harmful}: When we provide the main agent with the debug tools directly (start session, breakpoint, inspect, etc.), without a subagent wrapper performance either remains flat or degrades significantly. We see negligible change for GPT-5 and Claude Haiku, but for Claude Sonnet we see a drop of $\sim$18\% over baseline. Manual inspection of the trajectories revealed that the main agent rarely leverages the debug tools, calling at least one debug tool in only 17 (9\%) of the instances. Interestingly, it resolves 14 ($\sim$82\%) of those instances, while in the baseline only 11 ($\sim$64\%) of those instances succeed. In the cases where it uses debug tools, the subsequences corresponding to debugging are very short ($\leq$4 tool calls) compared to the Debug2Fix trajectories, where the agent can go into deeper debugging sessions to investigate complex questions posed by the main agent. Furthermore, the subagent doesn't get confused about tool orchestration because it has limited tool selection and has the sole purpose of debugging, unlike the main agent when presented with debug tools alongside its usual tool-set and is used for a wide range of tasks. 

\textbf{The subagent alone is often insufficient without tool-limiting} Adding the Debug Subagent as a tool and updating the prompt to let the model know of its presence yields varying call rates across models. Based on our results on GitBug-Java (Table ~\ref{tab:consolidated_results}), GPT-5 seems to be the most suggestible when it comes to adding a new tool for the agent to call, while Claude Sonnet 4.5 seems the most unwilling. This is unfortunate given how its the stronger model of the group and would result in even better overall performance should it leverage the tools. For GitBug-Java, by enforcing debugging before editing, our approach was able to raise the call rates to $\sim$99\% for all models and produce performance gains across the models.

\textbf{New tool's call-rate can vary by language} Directly comparing call rates between GitBug-Java and SWE-Bench Live for all our models, we can see how the willingness of models to adopt a Debug Subagent changes by language. In Java, all three models show a similar call rate by default (60-70\%), as seen in the Debug2Fix configuration. However, for Python the call rates diverge dramatically between the models. GPT-5 maintains it's $>$60\% call rate, but Haiku and Sonnet the call rates drop considerably. We hypothesize that this may be because certain models are more confident in their ability to fix Python issues statically or via print-debugging without needing a debugger, which is what they were likely trained to do. However, this reluctance to adopt the Debug Subagent also reflects the smaller improvements seen in their performance (Table~\ref{tab:consolidated_results_py}).

\textbf{Computational overhead to adding such a subagent} A natural concern with adding a new subagent with its own context and trajectory is the increased cost and added latency, not including the cost of running the debugger itself. LLM tokens and agent steps directly translate to cost and latency measures. Table~\ref{tab:consolidated_results} shows the breakdown of steps and tokens used by the main and subagent for GitBug-Java benchmark. This is so we can measure the change in the main agent's step count and token usage when it uses the subagent. For Claude Sonnet 4.5, we can see that we actually reduce the main agents token usage and steps when using our best configuration, while the subagent adds on average 33 steps and 400k tokens. For GPT-5 and Claude Haiku 4.5, the total tokens increases modestly. Examining the steps taken by each model, we can see that they also go up in most cases except for Claude Sonnet.

This variability in tool adoption across models and languages presents a challenge for agent developers. We encourage model providers to improve instruction-following for novel tools so that these systems can be made extensible. Models that selectively ignore available tools would continue to demonstrate poor performance when faced with real-world problem solving.

\textbf{Better tooling closes the gap between models} An interesting consequence of us incorporating a Debug Subagent is that, Debug2Fix allows weaker models to match or even exceed the performance of stronger models' baseline performance. As one can see for GitBug-Java, \textbf{GPT-5 with Debug2Fix scores (73.1\%) nearly the same as baseline Sonnet (75.7\%)}, despite the 15\% gap in their respective baselines. Similarly, \textbf{Debug2Fix helps Claude Haiku 4.5 outperform baseline Sonnet 4.5 by 6.6\%}. This suggests that equipping agents with better tooling can be just as, if not more impactful than simply upgrading to a more capable model, which has been the belief in Coding Agent community. This finding has implications for cost-sensitive areas where smaller or cheaper models with better tooling may be preferably to an expensive larger model.

\begin{table*}[h]
\centering
\footnotesize
\caption{Categorization of the kinds of questions we observed the main agent posing to the Debug Subagent in a sample of 50 instances.}
\label{tab:question_taxonomy}
\begin{tabular}{p{2.9cm}p{6.4cm}rr}
\toprule
\textbf{Category} & \textbf{Description} & \textbf{\% Cases} & \textbf{\# Instances} \\
\midrule
Exception Diagnosis & Identifying the type, origin, or cause of runtime exceptions. & 27.8\% & 17 \\
\midrule
Root Cause Analysis & Understanding why a specific behavior or bug occurs. & 25.3\% & 15 \\
\midrule
Local Variable Inspection & Querying the value of specific variables at a location. & 15.2\% & 11 \\
\midrule
Attribute Value Inspection & Inspecting object attributes or map entries. & 15.2\% & 9 \\
\midrule
Assertion Failure & Identifying which assertion fails and the actual vs expected values. & 6.3\% & 5
\\
\midrule
Code Reachability & Checking if execution reaches a specific location or branch. & 5.1\% & 4 \\
\midrule
Post-Fix Verification & Confirming that a code change resolves the issue. & 5.1\% & 3 \\
\bottomrule
\end{tabular}
\end{table*}

\begin{table*}[h]
\centering
\footnotesize
\caption{Failure modes observed in failed Debug2Fix instances despite having the Debug Subagent.}
\label{tab:failure_modes}
\begin{tabular}{p{3.8cm}p{7.5cm}rr}
\toprule
\textbf{Failure Mode} & \textbf{Description} & \textbf{\% Cases} & \textbf{\# Instances} \\
\midrule
Debugger Session Failed & Debug Subagent could not attach JDB due to failed build/test. This could be due to a variety of reasons like missing Gradle task, test process exited before debugger attached, etc. & 36\% & 18 \\
\midrule
Wrong Fix Despite Correct Debugging & Debug Subagent successfully answered the runtime question, but the main agent
applied an incorrect or incomplete fix. & 34\% & 17 \\
\midrule
High Complexity Bug & Bug required $>$3 debug sessions, yet the agent could not converge on the correct fix despite these multiple attempts. & 16\% & 8 \\
\midrule
Subagent API Error & Debug Subagent request failed due to server errors, which results in an empty response. &
8\% & 4 \\
\midrule
Subagent Static Analysis & Debug Subagent ends up doing static code analysis, providing incomplete information without actual runtime values. & 6\% & 3 \\
\bottomrule
\end{tabular}
\end{table*}

\subsection{Qualitative Analysis}
In addition to our Ablation Analysis, we conduct a qualitative analysis over the trajectories generated during our runs to better understand how agents leverage the Debug Subagent in practice. We first examine the kinds of runtime questions the main agent asks the Debug Subagent and whether this is in accordance with how it was instructed to use the subagent. Second, we identify common failure modes by inspecting the cases where the Debug Subagent was invoked, but the agent failed anyway. For all these analyses, we use the trajectories generated during the GPT-5 run over GitBug-Java under the Debug2Fix (w/ Tool Limit) configuration.

\subsubsection{Debug Subagent Usage Patterns}
We randomly sample 50 trajectories over GitBug-Java benchmark where the Debug Subagent was invoked by the main agent. We then look at all the questions asked by the main agent and manually classify them into 7 distinct categories shown in Table~\ref{tab:question_taxonomy}. For each category, we show how many instances had those categories of questions and the percentage of questions that fell under that category. Note that a given benchmark instance may ask multiple questions falling in different categories.

The most common use case is notably Exception Diagnosis, which covers $\sim$28\% of the questions asked. This is where the agent asks the subagent to identify which exception was thrown and where in code it originates. The second most common category is Root Cause Analysis, where the agent asks prying questions about the underlying behavior behind the bug. Local Variable Inspection and Attribute Value Inspection make up $\sim$30\% of the cases, where the main agent asks the subagent to observe runtime state and get values of variables and expressions. These categories mirror the core use cases developers use debuggers for in the real life and also what the main agent was asked to use the Debug Subagent for (Figure~\ref{fig:prompt_diff}).
\begin{figure}
    \centering
    \includegraphics[width=1\linewidth]{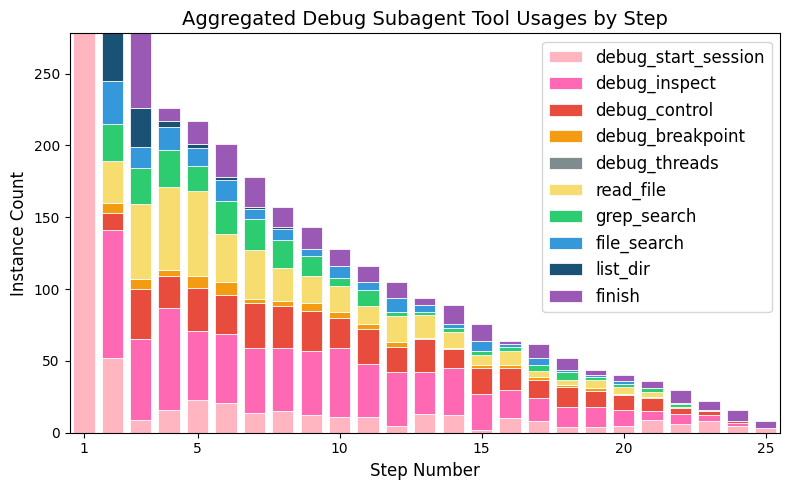}
    \caption{Plot showing an aggregated view of all the trajectories taken by the Debug Subagent. Each step shows a distribution of tools called within it. We can see the plot tapering to the right because more trajectories resolves as the subagent takes more steps.}
    \label{fig:debug_traj}
\end{figure}

\subsubsection{Typical Debug Subagent Workflow}
Figure~\ref{fig:debug_traj} shows the distribution of debug tool calls across step positions within the trajectories taken by the subagent. The plot shows a pattern to how the subagent orchestrates its debugging sessions. We can see that the first step is always Debug Start Session. This is by design of the subagent, where we execute the start session tool that atomically starts the program, attaches JDB / PDB and sets the initial breakpoints. We can see that the subsequent steps are dominated by Debug Inspect and Debug Control calls as the agent alternates between examining variables or expression values and stepping through execution. We also see the subagent reading files, grepping and setting breakpoints, which becomes less frequent in the later portion of the plot. 

The plot itself tapers to the right as more trajectories resolve and terminate, as seen via the Finish calls throughout the plot. Most trajectories finish within 10-15 steps, though many still go on until Debug Subagent trajectory limit of 25 steps.

\subsubsection{Common Debug2Fix Failure Modes}
To understand the limitations of our approach, we manually analyze a sample of 50 failed instances in GitBug-Java run, where the agent invoked the Debug Subagent, but still failed to resolve the bug. Table~\ref{tab:failure_modes} shows a categorization of the different failure modes we observed in the data. 

We can see that a big chunk of failures stem from infrastructure limitations and issues with the main agent itself rather than the Debug Subagent. We observe that the most prevalent failure mode is due to the debug session failing, which manifests as Debug Start Session tool call timing out in the trajectory. This is caused by the Debug Subagent not being able to attach to the test process due to some build failure. This highlights how buildability is a prerequisite for our approach i.e. if the project doesn't compile, the Debug Subagent won't be able to provide runtime insights. The second largest category is the main agent implementing the wrong fix despite correct answer from Debug Subagent. Following these, we have issues having to do with bugs requiring more than 3 debug sessions, the request failing or the subagent resorting to static analysis in some cases. These findings suggest that there may be potential improvements we can make to the underlying build logic behind the subagent, add retry logic as well as improve the translation of Debug Subagent diagnostics into fixes from the main agent. We leave these explorations to future work.

\section{Limitations}
While our study provides a valuable framework for agents to follow, we would like to highlight several limitations that present opportunities for future explorations.

\textbf{Language \& Benchmark Coverage} While our evaluation shows that our approach can we successfully applied to Python and Java, we have not tried this approach to other languages with mature ecosystems like C, C++, C\#, Rust, etc. Furthermore, we only tried one benchmark for both languages studied. In future work, we would like to explore more languages and benchmarks like SWE-Bench Pro~\cite{swepro} and Multi-SWE-Bench~\cite{zan2025multiswebenchmultilingualbenchmarkissue} with greater language coverage.

\textbf{Project Buildability} Our approach seems to rely on whether the project can be built and has executable tests. As seen in our Qualitative Analysis, 36\% of our failed instances stem from this issue. While this may be a reasonable constraint because developers do ultimately need to build and run the projects to do debugging themselves, this limits us to project with working build systems.

\textbf{Using Different Models} We use the same model for both the main and subagent across all our experiments. However, alternate configurations such as pairing different models for main agent and subagent may provide even better results. Potentially, one could even finetune a smaller model specialized for debugging, which may reduce cost of running such an agent. We leave these explorations to future work.

\textbf{Tool-Limiting Trade-off} While our tool limiting strategy worked for one of our benchmarks, it enforces a very rigid workflow on these models. This may lead to suboptimal performance or over-complication of simpler bugs that can be fixed without runtime information. A more adaptive strategy that selectively decides when to disable editing before debugging is needed to adapt this approach to real world agents.

\textbf{Debugging Overhead} While we measure the token usage and steps taken by the subagent, we don't measure the computational cost of running the debugger itself. It may not always be possible to run one depending on the environment.

\section{Conclusion}
In this work, we presented Debug2Fix, a framework that incorporates interactive debugging capabilities into coding agents through a specialized Debug Subagent. Our approach abstracts low-level debugger commands behind a unified interface that can be leveraged by the main agent without having to use debugger commands itself. Through extensive evaluation consisting of an ablation study and qualitative analysis of the trajectories, we show that Debug2Fix yields substantial improvements across all evaluated models and languages we study. Notably, we saw GPT-5 achieving a 21.8\% relative improvement over baseline, Claude Haiku 4.5 improving by 15.9\% and Claude Sonnet 4.5 improving by 12.9\%. 

One of the key findings of our approach is that better tooling can close the gap between models. With Debug2Fix, GPT-5 achieves nearly the same pass rate as baseline Claude Sonnet 4.5 on the same benchmark. Similarly, Claude Haiku 4.5 is able to surpass baseline Claude Sonnet 4.5 results when used with Debug2Fix. Through an ablation study we find that exposing debug tools directly to the main agent is ineffective and can even degrade performance i.e. subagent architecture is necessary. We also see models exhibiting varying degrees of reluctance in adopting new tools.

Debug2Fix represents a strong step towards coding agents that are on par with expert developers for tasks like bug fixing, by mirroring their workflows of using a debugger for tasks where static analysis proves insufficient. We think that this work will allow LLM-based coding agents to take on increasingly complex tasks. Finally, as a call to action, we implore LLM providers to support better instruction following, so that LLMs leverage novel tools more readily to enable more extensible and capable agent systems.

\textbf{Data Availability Statement:}
All datasets used in this work are publicly available benchmarks. The source code used to produce the results cannot be released due to our company policy. To support reproducibility, our paper includes the exact prompts, model configurations, and/or methodological details necessary to implement the proposed approach.

\bibliographystyle{IEEEtran}
\bibliography{iclr2024_conference}

\end{document}